\documentclass[aps,prb,twocolumn,floatfix]{revtex4}
\usepackage{tabularx,graphicx}
\begin{document}
\title{ Many body effects on  $c$- axis properties:
out of plane coherence and  bilayer splitting.\\
}
\author{ F. Guinea$^{\dag}$, M. P. L\'opez-Sancho$^{\dag}$,
and M.A.H. Vozmediano$^{\ddag}$.}
\affiliation{
$^{\dag}${I}nstituto de Ciencia de Materiales de Madrid,
Consejo Superior de Investigaciones Cient{\'{\i}}ficas,
 Cantoblanco, E-28049 Madrid, Spain.\\
$^{\ddag}$ Departamento de Matem\'aticas and Unidad Asociada of CSIC,
Universidad Carlos III de Madrid, Avda. de la Universidad 30,28911 Legan\'es,
Madrid, Spain.}
\date{\today}
\begin{abstract}
The out-of-plane hopping in layered metals with strong
electronic correlation is analyzed theoretically. By studying
the effects of in-plane interactions on the interlayer tunnelling
we investigate one of the oldest unresolved problems of
high-T$_c$ cuprates, the so-called bilayer splitting.

The strong electron-electron interactions within each layer reduce
the quasiparticle weight, and modify the hopping between layers.
We analyze the effect of the in-plane correlations on the interlayer
hopping using
a Renormalization Group scheme already applied to the problem of
interlayer coherence . The bilayer band splitting
 acquires a significant temperature and doping dependence, and
can be completely suppressed when the Fermi energy coincides with
a Van Hove singularity.

\end{abstract}
\maketitle
%
\section{Introduction}
The strong anisotropy and anomalous {$\bf{c}$}- axis
properties of the high-temperature cuprate superconductors
are among the most relevant features of these layered
materials and have greatly conditioned the theoretical
scenario \cite {andbook}. Although  intensively investigated,
the out-of-plane behaviour of the cuprates is not yet
well understood. The role of the multiple layers and
its relevance in the critical temperature $T_c$ and optimal doping
values of different compounds,  is not settled yet \cite {damascr}.

One of the most interesting issues has been the interaction
between $CuO_2$ planes in materials with two planes per unit cell.
As predicted by early band structure calculations, the interaction
between planes would split their electronic structure in bonding
and antibonding bands, the so-called bilayer band splitting.
This splitting was expected to be maximum at $(\pi,0)$ and vanish
along $(0,0)-(\pi,\pi)$, by symmetry effects \cite {chakra}.
The absence of the BS on  early angle-resolved photoemission  experiments
on $Bi_2Sr_2CaCu_2O_{8+\delta}$ (Bi2212)
aroused great expectation \cite {ding1}, in conection with the
interlayer pair tunneling model \cite {andbook}.
Recently, angle-resolved photoemission spectroscopy (ARPES) data
on bilayer cuprate Bi2212 have reported a bilayer band splitting
(BS) between bonding and antibonding bands in the
Brillouin zone (BZ) antinodal
region of overdoped samples, above and below $T_c$ \cite {feng,chuan}.
The bilayer energy splitting, due to the intrabilayer coupling, is of about
100 meV in the $(\pi, 0)$ region in the normal state and of about
20 meV in the superconducting state. The momentum and energy behavior of
the BS agree qualitatively with the bilayer Hubbard model calculations
\cite {liecht}. The report of BS as well  in underdoped and optimally doped
Bi2212 samples, not resolved before, by ARPES measurements
with different photon energies and high momentum and energy
resolution \cite {boris}, should be
seriously considered and put constrains in existing theories
and data analysis \cite {fengr,kordyuk,kim}.
The existence of BS in all doping ranges seems to confront with the
coherence issue. In this context, a crossover in the phase diagram
between the low temperature, overdoped side with coherent electronic
excitations, and the high temperature, underdoped side where this coherence
is lost, has been reported from ARPES data and resistivity measurements
of Bi2212  \cite {kamins}, in good agreement with
theoretical predictions \cite {us}, and with optical conductivity results \cite {taken}
from which the coherent-to-incoherent crossover as a function of doping
has been reported. In the overdoped region, conventional Fermi liquid behavior
is expected while in the underdoped region correlation effects would
deviate from the conventional physics yielding exotic behavior, in
agreement with the most common doping-temperature phase diagram for
the cuprates. A crossover from insulating-like, at high temperatures, to
metallic-like character at low temperatures in the direction perpendicular
to the layers, while being metallic within the layers, has been recently
reported by Valla {\it et al} from ARPES measurements in layered,
non-superconducting materials \cite {valla}, the change of effective dimensionality
being correlated with the existence or non-existence of coherence
within the layers.
The splitting of the $CuO_2$-plane states in the whole doping range
would imply the coherence of electronic states  in the ${\bf c}$ axis
at least to the intracell  distance between the two planes. At present,
consensus on the existence of BS has been reached only in the overdoped
regime, while in optimally doped and underdoped regime there is not
a clear agreement.
Recently \cite {chuang} high-resolution ARPES experiments on
Bi2212 covering the entire doping range, and
with fotons of 22 eV and 47 eV (ARPES matrix element effects for
fotons of 47eV strongly enhance the antibonding component of the
bilayer in the $(\pi, 0)$ region of the BZ) suggest that the size of the intracell
coupling is only weakly dependent on the doping level, therefore
some coherent coupling exists between the $CuO_2$ planes of the unit cell
even in  underdoped system. To explain the ${\bf c}$-axis transport
measurements where an insulating $\rho_c$ behavior is reported in
underdoped samples, ref \cite {chuang} argue that  intracell perpendicular hopping
will dominate the BS while the intercell perpendicular hopping will
control the ${\bf c}$-axis resistivity $\rho_c$

It is the purpose of this paper to analyze the role played by many-body effects on the
interlayer coupling of high-$T_c$ superconductors and its effects on the
bilayer splitting.

\section{Model and Method}
In spite of the great theoretical effort in high-$T_c$ superconductivity,
there is not yet a theory of consensus.
The Hubbard hamiltonian  is among the most studied models in the
hight-$T_c$ superconductors field, since it explains the main
low-energy physics results of the $CuO_2$ planes, which are
the common blocks to all families and where it is assumed to lie
the main physics of the cuprates. The insulating
parent compound is obtained at half-filling with antiferromagnetic
correlations, while upon doping the known phase diagram is obtained.
The stripe phase as well as d-wave superconductivity are well
described within the Hubbard model \cite {kim98}.
Furthermore, in the overdoped regime, where Fermi-liquid-like behavior seems
to hold, ARPES data of Bi2212 qualitatively agree with bilayer Hubbard model calculation
results for the  energy splitting \cite {feng}, a experimental
intrabilayer hopping $t_{\perp, exp} \approx 44 $ meV was obtained.

However, the anisotropy of the normal state transport properties:
electron motion in the {$\bf{c}$}-direction is incoherent in contrast with the
metallic behavior of the in-plane electrons as probed by the
different {$\bf \rho_{c}$} and {$\bf \rho_{ab}$}
resistivities\cite{fuji, ando}, cannot be understood in the
current theoretical framework.

The anomalous out of plane behavior of the cuprates has led
to the suggestion that conventional Fermi liquid theory fails
in these compounds\cite{andbook}. Both theoretical \cite {andbook,legg,TL01}
and  experimental \cite {marel} work has remarked the relevance of the
properties of the direction perpendicular  to the $CuO_2$ planes.

An alternative explanation of the
emergence of incoherent behavior in the out of plane direction has been
proposed in terms of the coupling of the interlayer electronic motion to
charge excitations of the system\cite{TL01}. This approach implicitly
assumes that electron electron interactions modify the in--plane
electron propagators in a non trivial way.

We will show that, even in the clean limit,
many body effects
can suppress the coherent contribution to the out
of plane electron hopping. The clean limit is defined
as that in which the length scale, $L$, over which
electrons remain coherent within the layers diverges.

The simplest formulation of the method replaces the
excitations of the system (such as electron-hole
pairs) by a bath of harmonic oscillators
with the same excitation spectrum.
This approach can be justified rigorously in one dimension,
and is always an accurate description of the response of
the system when the coupling of the quasiparticles to each individual
excitation is weak\cite{CL83}.

In the following,
we will assume a local interaction between electrons
close to the Fermi level, and the charge fluctuations of the system:
\begin{equation}
{\cal H}_{int} = c^\dag_i c_i \sum_{\bf \vec{k}}
V_i ( {\bf \vec{k}} ) \hat{\rho}_{\bf \vec{k}}
\label{int}
\end{equation}
where $c_i$  creates an electron at site $i$,
and $\hat{\rho}_{\bf \vec{k}}$ describes the charge
fluctuations of the environment, which are to be described
as a set of harmonic modes.
The Hamiltonian of the system is approximated as:
\begin{eqnarray}
{\cal H}_{e-b} &= &{\cal H}_{elec} + {\cal H}_{env} + {\cal H}_{int}
 \\
&= &\sum {\bf t}_{ij} c^\dag_i c_j + \sum \omega_k b^\dag_k
b_k + \sum g_{k , i} c^\dag_i c_i ( b^\dag_k + b_k ) \nonumber
\label{hamil}
\end{eqnarray}
where ${\cal H}_{elec}$ describes the individual quasiparticles,
${\cal H}_{env}$ stands for the set of harmonic oscillators
which describe the environment, and ${\cal H}_{int}$ defines
the (linear) coupling between the two.
The $b_k^\dag$
are boson creation operators,
the ${\bf t}_{ij}$ describe the electronic hopping processes, and the
information about the interaction between the electron in
state $i$ and the
system is defined by the the function\cite{CL83} $J_i ( \omega ) = \sum_k
| g_{k,i} |^2 \delta ( \omega - \omega_k )$.

Using second order perturbation theory and eq.(\ref{int}),
we can write\cite{CL83,SET}:
\begin{equation}
J_i ( \omega ) = \sum_{\bf \vec{k}} V_i^2 ( {\bf \vec{k}} )
{\rm Im} \chi ( {\bf \vec{k}} , \omega )
\end{equation}
where $\chi ( {\bf \vec{k}} , \omega )$ is the Fourier transform
of the density-density response of the system,
$\langle \hat{\rho}_{\bf \vec{k}} ( t ) \hat{\rho}_{\bf - \vec{k}}
( 0 ) \rangle$. The interaction in eq.(\ref{int}) is spin
independent. Other, more complicated, couplings can also
be taken into account, provided that the appropriate
response function is used.

The influence of the electron-boson coupling on the
electron propagators can be calculated to all orders if
the state $i$ is localized, that is, neglecting the
hopping terms in eq.({\ref{hamil}).  We find:
\begin{eqnarray}
\langle c^\dag_i ( t ) c_i ( t' )
\rangle
&\sim  &\langle c^\dag_i ( t ) c_i ( t' ) \rangle_0 \times
\nonumber \\ & &\exp \left\{ -
\int d \omega  \left[ 1 - e^{i \omega ( t - t' )} \right]
\frac{J_{i} ( \omega )}
{ \omega^2} \right\}
\label{Green}
\end{eqnarray}
where $\langle c^\dag_i ( t ) c_i ( t' ) \rangle_0 \sim
e^{i \varepsilon_i ( t - t')}$ is the Green's function in
the absence of the interaction. The method that we use
assumes that eq.(\ref{Green}) also holds in a system
with extended states. For a standard metallic system,
we must insert $\langle c^\dag_i ( t ) c_i ( t' ) \rangle_0 \sim
1 / ( t - t' )$ in eq.(\ref{Green}). It can be shown that
this approximation is exact at short times,
$W \ll ( t - t' )^{-1} \ll \Lambda$, where $W$ is an energy scale
related to the  dynamics of the electrons, and $\Lambda$ is
the upper cutoff in the spectrum of the environment.

The time dependence in eq.(\ref{Green}) is determined
by $\lim_{\omega \rightarrow 0} \chi ( {\bf \vec{k}} , \omega )$.
In a gapless, metallic system, we have
$\chi ( {\bf \vec{k}} , \omega ) \sim \alpha  ({\bf \vec{k}} )| \omega |$.
This
behavior, when inserted in eq.(\ref{Green}), leads to:
\begin{equation}
\langle c^\dag_i ( t ) c_i ( t' )
\rangle
\sim \frac{1}{( t - t')^{(1 + \alpha )}}
\label{green}
\end{equation}
where:
\begin{equation}
\alpha = \int_{| {\bf \vec{k}} | \ll L^{-1}}
d {\bf \vec{k}} V^2  ( {\bf \vec{k}} )
\alpha  ({\bf \vec{k}} )
\label{alpha}
\end{equation}
where $L$ is the scale of the region where the tunneling
process takes place. The value of $L$ is limited  by the
length over which the phase of the electronic wavefunctions
within the layers is well defined.
We assume that, in a translationally invariant system,
there is no dependence on the position of the local orbital, $i$.
This result implies that the frequency dependence of
the Green's function, in a
continuum description, can be written as:
\begin{equation}
\lim_{|{\bf \vec{r}} - {\bf \vec{r}}' | \rightarrow 0}
G ( {\bf \vec{r}} - {\bf \vec{r}}', \omega )
\propto | \omega |^{ \alpha}
\label{green_w}
\end{equation}
We can now use eq.(\ref{green}) to analyze the
interlayer tunneling by applying Renormalization Group methods.
The simplest case where this procedure has been used is for
the problem
of an electron tunneling between two
states, $i$ and $j$, which has been intensively
studied\cite{Letal87,W93}.  We integrate out
the high energy bosons, with energies
$\Lambda - d \Lambda \le
\omega_k \le \Lambda$ and rescaled
hopping terms are defined. As mentioned earlier,
eq.(\ref{green}) is valid for this range of energies.
The renormalization of
the hoppings is such that the properties of the
effective Hamiltonian at energies $\omega \ll \Lambda$
remain invariant. If the hoppings ${\bf t}_{ij}$ are small, any
physical quantity which depends on them can be
expanded, using time dependent perturbation theory, in powers of:
\begin{equation}
{\bf t}_{ij}^2 \langle c_i^\dag ( t ) c_j ( t ) c_j^\dag ( t' ) c_i ( t' )
\rangle \approx {\bf t}_{ij}^2 \langle c_i^\dag ( t ) c_i ( t' ) \rangle
\langle c_j ( t ) c_j^\dag ( t' ) \rangle
\label{perturbation}
\end{equation}
The integration of the high energy modes implies that the terms
in eq.(\ref{perturbation}) are restricted to $t \le \Lambda^{-1}$,
or, alternatively, the time  unit have to be rescaled\cite{C81},
$\tau' = \tau e^{d\Lambda / \Lambda }$, where $\tau \sim
\Lambda^{-1}$.
Using eq.(\ref{green}), the condition of keeping the
perturbation expansion in powers of the terms in eq.(\ref{perturbation})
invariant implies that:
\begin{equation}
{\bf t}_{ij}^2 \rightarrow {\bf t}_{ij}^2  e^{
\frac{d \Lambda}{\Lambda} \left( 2 + 2 \alpha \right)}
\end{equation}
which can also be used to define the scaling dimension of
the hopping terms. Finally,
\begin{equation}
\frac{\partial ( {\bf t}_{ij} / \Lambda )}{\partial l} =
- \alpha \frac{ {\bf t}_{ij} }{\Lambda}
\label{renor}
\end{equation}
where $l = \log ( \Lambda_0 / \Lambda )$, and $\Lambda_0$
is the initial value of the cutoff.

This approach has been successfully used to describe inelastic
tunneling in different situations in \cite{TL01,W90,SET,KF92,1D,SNW95,RG01}.

The analysis which leads to eq.(\ref{renor}) can be generalized to
study hopping between extended states, provided that we
can estimate the long time behavior of the Green's
function, $G ( {\bf \vec{k}} , t - t' )
= \langle c^\dag_{\bf \vec{k}} ( t ) c_{\bf \vec{k}} ( t' )
\rangle$.  This function is related to the local Green's function,
which is written in eq.(\ref{green_w}), by:
\begin{equation}
\lim_{|{\bf \vec{r}} - {\bf \vec{r}}' | \rightarrow 0}
G ( {\bf \vec{r}} - {\bf \vec{r}}', \omega ) =
\int d^D {\bf \vec{k}} G ( {\bf \vec{k}} , \omega )
\label{integral}
\end{equation}
where $D$ is the spatial dimension.
In the cases discussed below, the interaction is instantaneous in time,
and the non interacting Green's function can be written as:
\begin{equation}
G_0 ( {\bf \vec{k}} , \omega ) \propto \frac{1}{\omega}
{\cal F} \left( \frac{k_i^z}{\omega} \right)
\label{scaling_g0}
\end{equation}
where $z=1,2$.
In the following, we
assume that the interacting Green's function has the
same scaling properties, with the factor $\omega^{-1}$
replaced by $\omega^{-\delta}$ in eq.(\ref{scaling_g0}), where $\delta$
depends on the interactions.
This can be shown to be correct
in perturbation theory to all orders, in the models
studied below, because the corrections depend logarithmically
on $\omega$ (it is a well known fact for the
Luttinger liquid). Then, using eqs.
(\ref{green_w}), (\ref{integral})
and (\ref{scaling_g0}), we obtain:
\begin{equation}
G ( {\bf \vec{k}} , \omega ) \propto
| \omega |^{\alpha - D/z} {\cal F} \left( \frac{k_i^z}{\omega} \right)
\label{green_k}
\end{equation}
and $ {\cal F} ( u )$ is finite.
Thus, from the knowledge of
the real space Green's function, using
eq.(\ref{Green}),  we obtain $\alpha$, which, in turn, determines
the exponent $\alpha + D / z $ which characterizes
$G ( {\bf \vec{k}} , \omega )$. Generically, we can write:
\begin{equation}
G_{l,e} ( \omega ) \sim | \omega |^{\delta_{l,e}}
\label{scaling_green}
\end{equation}
where the subindices $l , e$ stand for localized and extended
wavefunctions. In terms of these exponents, we can generalize
eq.(\ref{renor}) to tunneling between general states to:
\begin{equation}
\frac{\partial ( {\bf t}_{ij}^{l,e} / \Lambda )}{\partial l} =
- \delta_{l,e} \frac{ {\bf t}_{ij} }{\Lambda}
\label{renor2}
\end{equation}
Before proceeding to calculations of $\delta_l$
and $\delta_e$ for various models,
it is interesting to note that, in general,
the response function of an electron gas in dimension $D > 1$ behaves as
$\lim_{\omega \rightarrow 0 , | {\bf \vec{k}} | \rightarrow 0}
\chi( {\bf \vec{k}} , \omega )
 \sim | \omega | / | {\bf \vec{k}} |$, so that, from
eq.(\ref{alpha}), $\lim_{L \rightarrow
\infty} \alpha \sim L^{(1-D)}$.
Thus, for $D > 1$, the contribution of the inelastic processes
to the renormalization of the tunneling vanishes for delocalized
states, $L \rightarrow \infty$.

%

\subsection{Van Hove singularities in the density of states.}

The dispersion relation of two-dimensional electronic systems
with a square lattice, i.e. the $CuO_2$ planes of the high-$T_c$
superconductors, usually present Van Hove singularities.
The Fermi surface of most hole-doped cuprates is close to a Van Hove
singularity.  The possible relevance of this fact to the
superconducting transition as well as to the anomalous behavior of the
normal state was put forward in the early times of the cuprates
and gave rise to the so called Van Hove scenario\cite{vhscenario}.
We will assume that the metallic layers are well described by
electrons in a square lattice, and that the Fermi level
is close to the $( \pi , 0 ) (A)$ and $( 0 ,  \pi ) (B)$
points of the Brillouin Zone (BZ). Close to these points,
the dispersion relation can be parametrized as:
\begin{equation}
\varepsilon_{A,B} ({\vec {\bf k}} ) \approx \frac{k_x^2}{2 m_{x,y}}
\mp \frac{k_y^2}{2 m_{y,x}}
\label{disp}
\end{equation}
where $m_x$ and $m_y$ are parameters which can be estimated from
the band structure of the model.
In the following, we will consider the renormalization
of the interlayer tunneling associated to these
regions in the BZ.

The response function at low energies
and small wavevectors has been  computed in \cite{GGV96}:
\begin{equation}
{\rm Im} \: \chi ({\vec{\bf k}}, \omega) =  \sum_{i = A,B}
\frac{1}{4 \pi\varepsilon_i ({\vec{\bf k}})} \left(\left|
\omega+\varepsilon_i
({\vec{\bf k}})
\right| - \left| \omega-\varepsilon_i ({\vec{\bf k}}) \right|\right)\;,
\end{equation}
where $\varepsilon_i({\vec{\bf k}})$ is the dispersion relation (\ref{disp}).

The long time dependence of the Green's function is determined
by the low energy behavior of $\chi$:
$\lim_{\omega \to 0}{\rm Im} \: \chi ({\vec{\bf k}}, \omega) \sim
\sum_{i = A,B} | \omega | / \varepsilon_i ({\vec{\bf k}})
$.
We assume that the interaction between
the electrons and the density fluctuations is
short ranged
as before.
The divergence of ${\rm Im} \chi$
when $\varepsilon_{A,B} = 0$ implies that the integral
in eq.(\ref{alpha}) diverges logarithmically
as $L \rightarrow \infty$,
as in the two previous cases, irrespective of
the details of the interaction, $V ( {\bf \vec{k}} )$.
Because of this divergence, it is convenient to
shift slightly the chemical potential $ \varepsilon_F$
away from the saddle point\cite{GGV96}, $\varepsilon_{VH}$.
A finite value of $| \varepsilon_F - \varepsilon_{VH} |$ implies the
existence of a length scale, $L_0 \sim
[ {\rm Max} ( m_x , m_y ) | \varepsilon_F - \varepsilon_{VH} | ]^{- 1/2}$, which
regularizes the ${\bf \vec{k}}$ integrals
in eq.(\ref{alpha}).

Using a local potential,
we find:
\begin{equation}
\alpha \sim ( U / W )^2
\log^2 ( L / L_0 )
\end{equation}
where $W$ is an energy scale of the order of the
width of the conduction band.
The dependence of $\alpha$ on $L$ goes
as $\log^2 ( L )$, as in other physical quantities
in this model\cite{GGV96}.

In the two dimensional model,  $z = 2$, so that
$\delta_l = \alpha$, as estimated above,
and $\delta_e = \alpha - 1$. The divergence of $\alpha$
implies that tunneling between localized and also between extended
states is suppressed at low temperatures.
In addition, the effective electron-electron coupling, $U$,
grows at low energies or temperatures,
until a scale at which the system
is unstable and a phase transition takes place\cite{GGV96}.
This effect enhances the suppression of
interlayer hopping.

\section{Bilayer band splitting}

To study de bilayer band splitting (BS) we apply  the method described
above and consider the interlayer tunnelling  between two $CuO_2$
planes within the Van Hove scenario. We will consider the electronic
states near the two inequivalent saddle points of the square lattice.
The lowest order corrections to the bare couplings
included in the {\bf {t-t'}}- Hubbard  Hamiltonian,
will give the anisotropic screening of the interactions \cite {GGV96},
by the Kohn-Luttinger mechanism,
in contrast to the isotropic three-dimensional metal \cite {kohnlu}.
The parameters of the model will be the interlayer hopping, the distance
of the Fermi level $\varepsilon_F$ to the Van Hove singularity $\varepsilon_{VH}$
given by the band structure, the band width $W$ and the strength of
the interaction $\alpha$. The effective Hamiltonian will be defined
at each scale of the couplings. The $| \varepsilon_{VH}-\varepsilon_F|$
defines the low energy cutoff while the bandwidth $W$ defines
the high-energy cutoff of the Renormalization Group approach.
The hopping renormalization gives the
$t_{eff} \approx t \left (  \frac {| \varepsilon_{VH}-\varepsilon_F|} {W} \right )^{ \alpha }$

We analyze the phase diagram relating the strength of the couplings
to the doping level.  The underdoped regime would correspond to
the strongly correlated system. The interacition $\alpha > 1$,
the Van Hove singularity lies at the Fermi level so that
the low-energy cutoff $|\varepsilon_{VH} -\varepsilon_F |=0$; no BS
occurs at this regime, a broad structureless background will
extend in the scale of $W$ as it is shown in Fig 1.
\begin{figure}
\resizebox{\columnwidth}{!}{\includegraphics{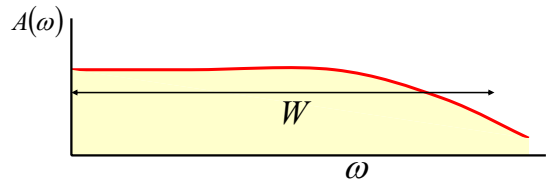}} \caption{
\label{figbsscoup} Schematic representation of spectral function
A($\omega$) versus frequency for the strong coupling case. The
band width W indicates the high-energy cutoff. }
\end{figure}

The system is instable at low energies,
and low temperature phase transition to the superconducting
phase are highly probable. The underdoped-optimally doped regime would correspond
as well to strongly correlated system, $\alpha > 1$, but in the frequency
region below the low-energy cutoff a renormalized bilayer splitting
will appear, since quasiparticles would be renormalized.
The renormalization factor
depends on the cutoffs as

$ Z= \left ( \frac {\left |\varepsilon_{VH} -\varepsilon_F \right|} {W} \right )^{ \alpha }$

Between the low and high energy cutoffs, a broad background
without any structure would be obtained, as schematically represented in Figure 2.

\begin{figure}
\resizebox{\columnwidth}{!}{\includegraphics{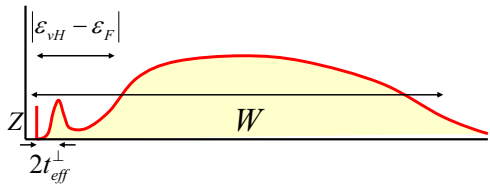}} \caption{
\label{figbsscou2} Same as Fig.1 with renormalized bilayer
splitting at frequencies below the low-energy cutoff, $
|\varepsilon_{VH} -\varepsilon_F |$. The band width W indicates
the high-energy cutoff. }
\end{figure}

The overdoped regime, as stated above, would correspond to the
weakly correlated case, $\alpha < 1$. Electron correlations are
screened and, as schematically represented in Figure 3, the renormalization
of the bilayer splitting would be weaker than in the preceding case.
An incoherent tail of the spectral weight is shown with a power law
decay at high energies, $A(\omega) \approx \frac {C} {\omega^{1- \alpha }}sin(\pi \alpha/2)$
always below the high-energy cutoff.

\begin{figure}
\resizebox{\columnwidth}{!}{\includegraphics{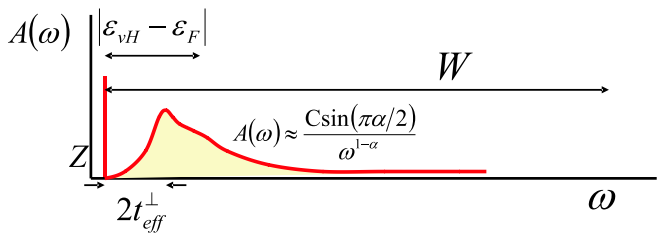}} \caption{
\label{figbswcou} Same as Fig.1 with  bilayer splitting at
frequencies below the low-energy cutoff, $ |\varepsilon_{VH}
-\varepsilon_F |$. The spectral weight decays at high energies.
The band width W indicates the high-energy cutoff. }
\end{figure}




\section{Conclusions.}
The effects of electron-electron correlations on the out-of-plane
properties of layered systems have been studied.
We have discussed the suppression of interlayer tunneling by
inelastic processes in two dimensional systems in the clean limit.
Our results suggest that, when perturbation theory
for the in--plane interactions leads to logarithmic
divergences,
the out of plane tunneling acquires
a non trivial energy dependence.
The conductance goes to zero as $T \rightarrow 0$
if the Fermi level
of the interacting electrons lies at a Van Hove
singularity.  Coherence would be suppressed and the bilayer splitting would vanish
as a consequence of the strong correlations.
When $|\epsilon_{VH} -\epsilon_F | \neq 0 $, a strongly renormalized
bilayer splitting occurs below $|\epsilon_{VH} -\epsilon_F |$ which
marks the low-energy cutoff of the system. These two situations
correspond to the underdoped and optimally doped regions in the
cuprate scenario. In the overdoped case, quasiparticles would be weakly
renormalized and the bilayer splitting will be present.

Thus, we have shown that insulating behavior in the
out of plane direction is not incompatible with
gapless
 or even superconducting
in--plane properties,
although the in plane properties are also
markedly different from those of an ordinary
Fermi liquid.

{\it Acknowledgments}.
The financial support of the CICyT (Spain), through
grant nº  MAT2002-04095-C02-01
is gratefully acknowledged.

\end{document}